\documentclass[twocolumn]{aastex63}
\usepackage{epsfig}
\usepackage{graphicx}
\usepackage{float}
\usepackage{amsmath}
\usepackage{color}
\usepackage{amssymb}
\usepackage{amsfonts}
\usepackage{units}
\usepackage{bm}
\usepackage{longtable}
\usepackage{booktabs}
\usepackage[mathscr]{euscript}

\def\be{\begin{eqnarray}}
\def\ee{\end{eqnarray}}

\shorttitle{Multiband Photometry of SN 2023ixf}
\shortauthors{Yang, Liu \& Pan et al.} 

\begin{document}

\title{Multiband Simultaneous Photometry of Type II SN 2023ixf with \emph{Mephisto} and the Twin 50-cm Telescopes}

\correspondingauthor{Yuan-Pei Yang (ypyang@ynu.edu.cn), Xiangkun Liu (liuxk@ynu.edu.cn), Xiaowei Liu (x.liu@ynu.edu.cn)} 

\author[0000-0001-6374-8313]{Yuan-Pei Yang}
\affiliation{South-Western Institute for Astronomy Research, Yunnan University, Kunming, Yunnan 650504, P.R. China}
\affiliation{These authors contributed equally to this work}

\author[0000-0003-0394-1298]{Xiangkun Liu}
\affiliation{South-Western Institute for Astronomy Research, Yunnan University, Kunming, Yunnan 650504, P.R. China}
\affiliation{These authors contributed equally to this work}

\author[0009-0002-7625-2653]{Yu Pan}
\affiliation{South-Western Institute for Astronomy Research, Yunnan University, Kunming, Yunnan 650504, P.R. China}

\author[0000-0002-8700-3671]{Xinzhong Er}
\affiliation{South-Western Institute for Astronomy Research, Yunnan University, Kunming, Yunnan 650504, P.R. China}

\author[0000-0002-0409-5719]{Dezi Liu}
\affiliation{South-Western Institute for Astronomy Research, Yunnan University, Kunming, Yunnan 650504, P.R. China}

\author[0009-0006-1010-1325]{Yuan Fang}
\affiliation{South-Western Institute for Astronomy Research, Yunnan University, Kunming, Yunnan 650504, P.R. China}

\author[0000-0002-8109-7152]{Guowang Du}
\affiliation{South-Western Institute for Astronomy Research, Yunnan University, Kunming, Yunnan 650504, P.R. China}

\author[0000-0002-7714-493X]{Yongzhi Cai} 
\affiliation{Yunnan Observatories, Chinese Academy of Sciences, Kunming 650216, P.R. China}
\affiliation{Key Laboratory for the Structure and Evolution of Celestial Objects, Chinese Academy of Sciences, Kunming 650216, P.R. China}
\affiliation{International Centre of Supernovae, Yunnan Key Laboratory, Kunming 650216, P.R. China}

\author[0009-0001-5333-8350]{Xian Xu}
\affiliation{South-Western Institute for Astronomy Research, Yunnan University, Kunming, Yunnan 650504, P.R. China}

\author[0009-0000-4068-1320]{Xinlei Chen}
\affiliation{South-Western Institute for Astronomy Research, Yunnan University, Kunming, Yunnan 650504, P.R. China}

\author[0009-0006-5847-9271]{Xingzhu Zou}
\affiliation{South-Western Institute for Astronomy Research, Yunnan University, Kunming, Yunnan 650504, P.R. China}

\author[0000-0001-5737-6445]{Helong Guo}
\affiliation{South-Western Institute for Astronomy Research, Yunnan University, Kunming, Yunnan 650504, P.R. China}

\author[0000-0001-5561-2010]{Chenxu Liu}
\affiliation{South-Western Institute for Astronomy Research, Yunnan University, Kunming, Yunnan 650504, P.R. China}

\author[0000-0001-8278-2955]{Yehao Cheng}
\affiliation{South-Western Institute for Astronomy Research, Yunnan University, Kunming, Yunnan 650504, P.R. China}

\author[0000-0001-7225-2475]{Brajesh Kumar}
\affiliation{South-Western Institute for Astronomy Research, Yunnan University, Kunming, Yunnan 650504, P.R. China}

\author[0000-0003-1295-2909]{Xiaowei Liu}
\affiliation{South-Western Institute for Astronomy Research, Yunnan University, Kunming, Yunnan 650504, P.R. China}

\begin{abstract}

SN 2023ixf, recently reported in the nearby galaxy M101 at a distance of $6.85~{\rm Mpc}$, was one of the closest and brightest core-collapse supernovae (CCSNe) in the last decade. In this work, we present multi-wavelength photometric observation of SN 2023ixf with the Multi-channel Photometric Survey Telescope (Mephisto) in $uvgr$ bands and with the twin 50-cm telescopes in $griz$ bands. 
We find that the bolometric luminosity reached the maximum value of $3\times10^{43}~{\rm erg~s^{-1}}$ at 3.9 days after the explosion and fully settled onto the radioactive tail at $\sim90$ days. The effective temperature decreased from $3.2\times10^4~{\rm K}$ at the first observation and approached to a constant of $\sim(3000-4000)~{\rm K}$ after the first two months.
The evolution of the photospheric radius is consistent with a homologous expansion with a velocity of $8700~{\rm km~s^{-1}}$ in the first two months, and it shrunk subsequently. 
Based on the radioactive tail, the initial nickel mass is about $M_{\rm Ni}\sim 0.098M_\odot$.
The explosion energy and the ejecta mass are estimated to be $E\simeq(1.0-5.7)\times10^{51}~{\rm erg}$ and $M_{\rm ej}\simeq(3.8-16)M_\odot$, respectively.
The peak bolometric luminosity is proposed to be contributed by the interaction between the ejecta and the circumstellar medium (CSM). We find a shocked CSM mass of $M_{\rm CSM}\sim0.013M_\odot$, a CSM density of $\rho_{\rm CSM}\sim2.5\times10^{-13}~{\rm g~cm^{-3}}$ and a mass loss rate of the progenitor of $\dot M\sim0.022M_\odot~{\rm yr^{-1}}$.

\end{abstract} 


\section{Introduction}

Core-collapse supernovae (CCSNe) arise from the core collapses of massive stars ($M\gtrsim8M_\odot$) in their final evolutionary stages when the radiation pressure from the nuclear fusion in the core cannot resist its own gravity. 
Given the extreme physical conditions, many details of the explosion process remain elusive.
Based on the observed spectra and light curves, CCSNe could be classified into types IIP, IIL, IIn, IIb, Ib, and Ic \citep{Filippenko1997, Modjaz2019}.
CCSNe retaining a large portion of the hydrogen envelope are known as Type II SNe, and their observed spectra are dominated by the Balmer lines.
CCSNe having no hydrogen envelope but showing helium features are known as Type Ib. Those that have lost much or all of their helium envelope are known as Type Ic. 

SNe II are among the most commonly observed SNe. They mark the deaths of massive stars and play an important role in the cosmic chemical enrichment. 
The light curves and spectra of SNe II give insight into the SN explosion mechanisms and the progenitor properties. The observed features of the light curves and the spectra of the SNe II vary across the sub-classes that depend on the fraction of the losses of the outer layer(s) \citep[e.g.,][]{Branch17}. 
SNe IIP have thick hydrogen envelopes and show long plateau phases ($\sim100$ days) with roughly constant or slowly declining luminosity \citep{Anderson2014a}, and about half of CCSNe are SNe IIP. SNe IIL have thinner envelopes and decline more quickly and account for 6\% of all CCSNe. The differentiation between SNe IIP and SNe IIL is mainly based on photometry, although it is not clear that the light curves of SNe IIL can be rigorously distinguished from those of SNe IIP \citep{Anderson2014a,Sanders15,Holoien16}.
SNe IIn, accounting for 10\% of all CCSNe, are characterized by narrow ($\sim10-100~{\rm km~s^{-1}}$) or intermediate-width ($\sim1000~{\rm km~s^{-1}}$ ) hydrogen emission lines attributed to the circumstellar interaction \citep{Schlegel1990}. It is noteworthy that some SNe IIL also show narrow emission lines. Meanwhile, some SNe IIn present linear-decline phases in their light curves \citep{Di02,Smith09}. It is unclear whether SNe IIL can be rigorously distinguished from SNe IIn.
Therefore, it is crucial to investigate the observational properties of Type II SNe in detail.

Very recently, SN 2023ixf was reported on 2023 May 19.727 UTC in the nearby face-on ``Pinwheel Galaxy'' Messier 101 (M101; NGC 5457) \citep{Itagaki23} at a distance of $6.85\pm0.15~{\rm Mpc}$ \citep{Riess22}, and it was one of the closest and brightest CCSNe in the last decade.
The evolution of SN 2023ixf has been extensively monitored by numerous facilities all over the world, and both professional and amateur astronomers reported their early observations of photometry  \citep[e.g.,][]{Balam23,Brothers23,Chen23,DAvanzo23,Daglas23,Desrosiers23,Fowler23,Gonzalez-Carballo23,Kendurkar23a,Kendurkar23b,Kendurkar23c,Koltenbah23,Li23,Maund23,Mayya23,Pessev23,Sgro23,Silva23a,Silva23b,Singh23,Vannini23a,Vannini23b,Vannini23c,Villafane23,Zimmerman23} and spectroscopy \citep[e.g.,][]{BenZvi23a,BenZvi23b,Lundquist23,Stritzinger23,Sutaria23a,Sutaria23b,Zhang23,Zimmerman23} of this object.
SN 2023ixf was classified as a Type II SN \citep{Perley23} and showed a strong blue continuum and prominent optical flash-ionization features (H, He, N, and C) in its early-phase spectrum, all indicative of the presence of circumstellar medium (CSM). 
Interestingly, the early photometric and spectroscopic studies have revealed evidence of the interaction between the SN ejecta and a dense confined CSM that boosted the early-phase optical luminosity of the SN \citep{Bostroem23,Hiramatsu23,Hosseinzadeh23,Jacobson-Galan23,Smith23,Teja23,Yamanaka23,Li23}. The X-ray detection, starting about four days after the explosion, also supported the scenario of the circumstellar interaction \citep{Chandra23,Grefenstette23,Mereminskiy23}. So far no evidence has been found for statistically significant emission in the sub-millimeter \citep{Berger23}, radio \citep[at 10 GHz][]{Matthews23a,Matthews23b,Matthews23c,Matthews23d}, gamma-rays \citep{Muller23} and neutrinos \citep{Thwaites23,Guetta23}.
Further, the spectropolarimetric investigation of this event revealed an aspherical SN explosion and distinct geometry of the CSM \citep{Vasylyev2023} and provided a temporally-resolved description of a massive-star explosion \citep{Zimmerman23}.

In addition to the SN 2023ixf itself, the currently available photometric information pointed to a luminous red supergiant with a dense shell of circumstellar material as its progenitor candidate \citep[][]{Jencson23,Kilpatrick23,Xiang23,Liu23}. The progenitor has been estimated to have a bolometric luminosity of $L\sim(10^{4.7}-10^{5.4})L_\odot$, an effective temperature of $T_{\rm eff}\sim(3200-3900)~{\rm K}$, and a mass of $M\sim(9-20)M_\odot$ \citep{Niu23,Hosseinzadeh23,Pledger23,Jencson23,Kilpatrick23,Soraisam23,Neustadt23,Xiang23}. The estimated values of the parameters depend on the model of the progenitor and its dust-driven wind environment reported in the literature, though \citet{VanDyk23} recently proposed that the progenitor candidate has a spectral energy distribution (SED) and luminosity strikingly similar to a Galactic red supergiant analog (IRC-10414).

In this work, we present and model multi-wavelength photometry of SN 2023ixf based on the observations with the \textbf{M}ulti-chann\textbf{e}l \textbf{Ph}otometr\textbf{i}c \textbf{S}urvey \textbf{T}elesc\textbf{o}pe (Mephisto) and with the twin 50-cm telescopes. We choose 2023 May 18 18:00 (MJD 60082.75) as phase zero, which is approximately midway between the first detection of the SN and the deep nondetection,
as proposed by \citet{Hosseinzadeh23}. All times in the results and figures are related to this phase zero.
The paper is organized as follows. 
We describe our photometric observations and present the observed properties of SN 2023ixf in Section \ref{observation}. We describe the evolution of the bolometric luminosity, effective temperature, and photospheric radius in Section \ref{fireball}.
We analyze the physical properties of SN 2023ixf in Section \ref{model}.
The results are discussed and summarized in Section \ref{conclusion}. The convention $Q_x=Q/10^x$ is adopted in cgs units unless otherwise specified.

\section{Multiband Photometric Observations and Data Reduction }\label{observation}

\begin{figure}
    \centering
    \includegraphics[width = 1.0\linewidth, trim = 0 0 0 0, clip]{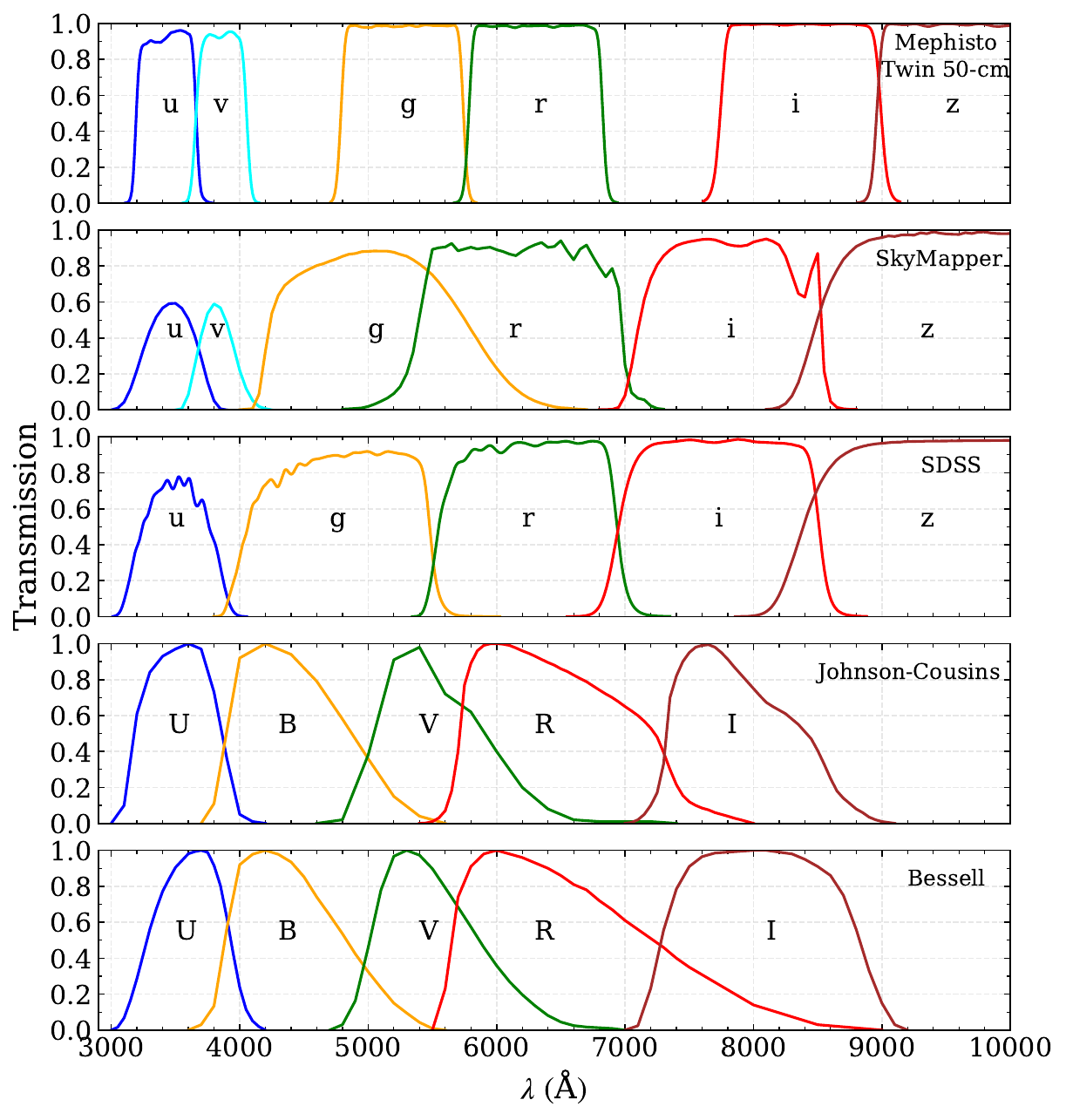}
    \caption{The transmission curves of the filters of Mephisto and the twin 50-cm telescopes compared with other optical telescopes and filter systems. The transmission curves of SkyMapper are given by \citet{Bessell11}, and the transmission curves of SDSS and the filter systems of Johnson-Cousins and Bessell are from \url{http://svo2.cab.inta-csic.es/theory/fps/}.}\label{transmission}
\end{figure}

Since the discovery alert of SN 2023ixf, we carried out continuous photometric follow-up observations from May 20, 2023 with the \textbf{M}ulti-chann\textbf{e}l \textbf{Ph}otometr\textbf{i}c \textbf{S}urvey \textbf{T}elesc\textbf{o}pe (Mephisto) under commissioning. The data presented in this work were collected between May 20, 2023 and February 24, 2024. Mephisto is a wide-field multi-channel telescope, the first of its type in the world, located at Lijiang Observatory (IAU code: 044) of Yunnan Astronomical Observatories, Chinese Academy of Sciences (CAS), with longitude $100^{\circ}01'48''$ East, latitude $26^{\circ}41'42''$ North and altitude $3200$m. The telescope has a $1.6$m primary mirror and covers a field-of-view (FOV) of $2^\circ$ in diameter. It is capable of imaging the same FOV in three optical bands simultaneously and delivering real-time, high-quality colors of unprecedented accuracy of surveyed celestial objects. Mephisto was equipped with two commercial Oxford Instruments/Andor Technology iKon-XXL single-chip CCD cameras for the blue and yellow channels during most of this period, allowing imaging respectively $uv$ and $gr$ filters. Each camera employs an e2v CCD231-C6 $6144\times 6160$ sensor with a pixel size of $15~\mu$m (corresponding to $0.429$ arcsec projected on the sky), and covers an area about a quarter of the full FOV. The wavelength coverages of the $uvgr$ filters are respectively $320-365$, $365-405$, $480-580$, and $580-680$ nm, with a corresponding average efficiency of $90\%$, $93\%$, $99\%$, and $98\%$, as shown in Figure \ref{transmission}. The observations were performed with filter combinations of $ug$ and $vr$ for the blue and yellow channels in a simultaneous mode. To ensure optimal signal-to-noise ratios of the target, the exposure times in different bands were adjusted accordingly over the time, ranging from 10 to 300s, and 2 to 180s, for the $uv$ and $gr$ bands, respectively. After eliminating data of bad quality, we are left with 242, 229, 327, and 317 exposures collected in 41 days in the $u$, $v$, $g$, and $r$ bands, respectively. 

As auxiliary photometric telescopes of Mephisto, the twin 50-cm telescopes, also located at Lijiang Observatory, were also involved in the observations of SN 2023ixf. The twin telescopes are of model Alluna RC20, with a flat-field corrector, and have a $505$mm optical aperture and a $f/8.1$ focal ratio. Each telescope is equatorial mounted with a 10 Micron GM4000 mount, and equipped with a  FLI ML 50100 CCD camera. The camera employs an On Semi KAF-50100 sensor, with $8176 \times 6132$ pixels of size  $6\mu$m (corresponding to $0.3$ arcsec projected on the sky). The observations were performed simultaneously with the twin telescopes, in two bands of filters $griz$. The CCDs were binned by 2 at the readout, resulting in a pixel scale of $0.6$ arcsec. The $gr$ filters have transmission curves similar to those of Mephisto, and the wavelength coverages of $iz$ filters are $775-900$  and $900-1050$ nm, respectively, with an average efficiency of $99\%$. In most cases, the exposure times were 10, 10, 60, and 120s in $griz$ bands, respectively. In total, 34, 47, 50, and 30 exposures were collected in 12 days in $g$, $r$, $i$, and $z$ bands, respectively, after excluding abnormal or overexposed images. 

\begin{figure}
    \centering
    \includegraphics[width = 1.0\linewidth, trim = 250 50 250 80, clip]{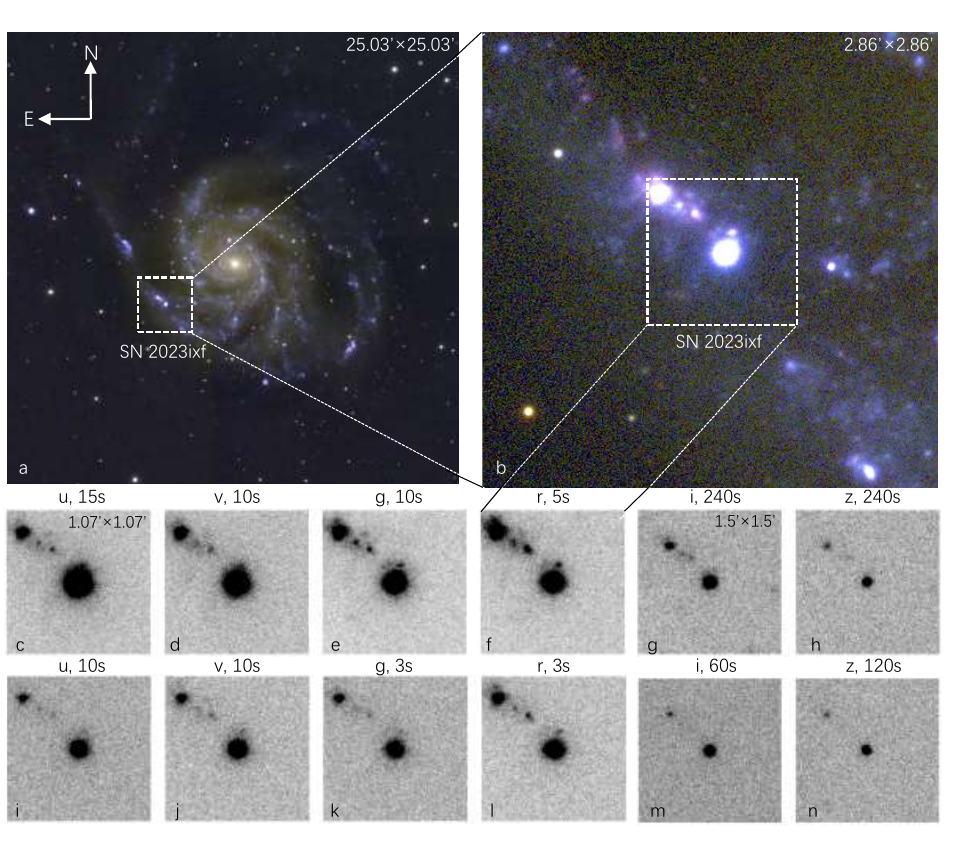}
    \caption{Images of SN 2023ixf with Mephisto (a-f,i-l) in $uvgr$ bands and the twin 50-cm telescopes (g,h,m,n) in $iz$ bands. Panel (a): the $vgr$-band composite image of the ``Pinwheel Galaxy'' Messier 101 with Mephisto on 2023 May 20. Panel (b): the image of the region near SN 2023ixf. Panel (c-h): images of SN 2023ixf in $uvgriz$ bands with different exposures on 2023 May 21. Panel (i-n): images of SN 2023ixf in $uvgriz$ bands with different exposures on 2023 June 4.}\label{2023ixf_img}
\end{figure}

We first performed pre-processing for each raw image, including bias/dark subtraction, flat-fielding with twilight flats, and gain correction. After pre-processing, the satellite trajectories and cosmic rays were removed. We then carried out astrometric calibration, using stars from the Gaia EDR3 catalog as the reference \citep{GaiaEDR3}. The astrometric calibration is typically accurate to 40 mas. Multiple observations of HST standard stars covering a sufficiently large airmass range were conducted on two photometric nights, April 9 and May 29, 2023. Those observations were used to derive the zero points of magnitude and first-order atmospheric extinction coefficients of the observed bands for absolute photometric calibration for data obtained with the Mephisto and the twin 50-cm telescopes. On this basis, we further performed relative photometric calibration to account for non-photometric observing conditions: 1) We first chose an image obtained in a photometric night with small airmass and good image quality with FWHM $<2$ arcsec as the reference image for flux alignment;
2) We scaled the flux levels of the remaining images to that of the reference image using stars of signal-to-noise ratios S/N $> 50$.
We then applied the following procedures to construct the multiband light curves of SN 2023ixf: 

(1) Reference and target images: High-quality images obtained before the supernova explosion were used as the reference images for image subtraction, while images exposed after the explosion were target images.

(2) Image alignment: For each target image, we used SWARP \citep{Bertin2002} to subtract the sky background and align it with the reference image for image subtraction for a selected area centered on SN 2023ixf. The selected areas were $3000\times 3000$, and $1500 \times 1500$ pixels, for images obtained with Mephisto and the twin 50-cm telescopes, respectively. Note the pixel sizes were 0.429 and 0.6 arcsec for the Mephisto and 50-cm image frames.

(3) Point Spread Function (PSF) model construction: we then constructed the PSF model for each image after alignment using SExtractor \citep{Bertin1996} and PSFex \citep{Bertin2011}, and acquired the PSF at the position of SN 2023ixf.

(4) Image subtraction and PSF photometry: After conducting mutual PSF convolution on both target images and reference images for image subtraction, we obtained the subtracted images by subtracting target images from reference image, and performed PSF photometry to construct the light curves.

In Figure \ref{2023ixf_img}, we plot the images of SN 2023ixf with Mephisto and the twin 50-cm telescopes in $uvgriz$ bands, the middle panels and the bottom panels correspond to the images of SN 2023ixf region on 2023 May 21 and 2023 June 4, respectively.
After obtaining the differential photometric measurements, we further corrected the magnitudes for the interstellar extinction. The foreground extinction from the Milky Way in the line of sight of SN 2023ixf is $E(B-V)_{\rm MW}$= 0.008\,mag \citep{Schlegel98,Schlafly11}. For the host galaxy extinction, we used the result of \citet{Liu23} obtained by analyzing the spectral data from the Hobby Eberly Telescope Dark Energy Experiment (HETDEX), who obtained an extinction value of $E(B-V)_{\rm host} = 0.06 \pm 0.14$ \,mag that is slightly larger than $E(B-V)=0.031~{\rm mag}$ given by some previous works \citep{Smith23,Teja23}.
Using the extinction law of \citet{Cardelli89} and assuming $R_V$ = 3.1, we obtain the extinction values of 0.338, 0.316, 0.221, 0.183, 0.117 and 0.095\,mag accounting for both the Milky Way and host galaxy for the $uvgriz$ bands, respectively. 

\begin{figure*}
    \centering
    \includegraphics[width = 1\linewidth, trim = 0 0 0 0, clip]{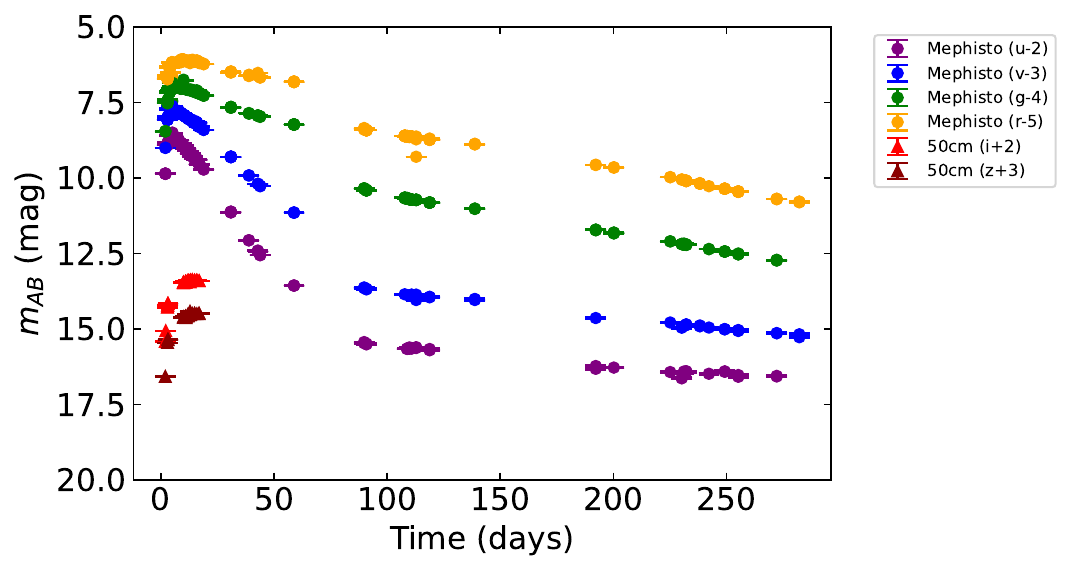}
    \caption{Multiband light curves of SN 2023ixf obtained with Mephisto and the twin 50-cm telescopes. The photometry data of Mephisto and the twin 50-cm telescopes is shown in the Appendix. Offsets have been added to the magnitudes of different bands for clarity, as specified in the legend.}\label{data_lc}
\end{figure*}

\begin{figure}
    \centering
    \includegraphics[width = 1.1\linewidth, trim = 0 0 0 0, clip]{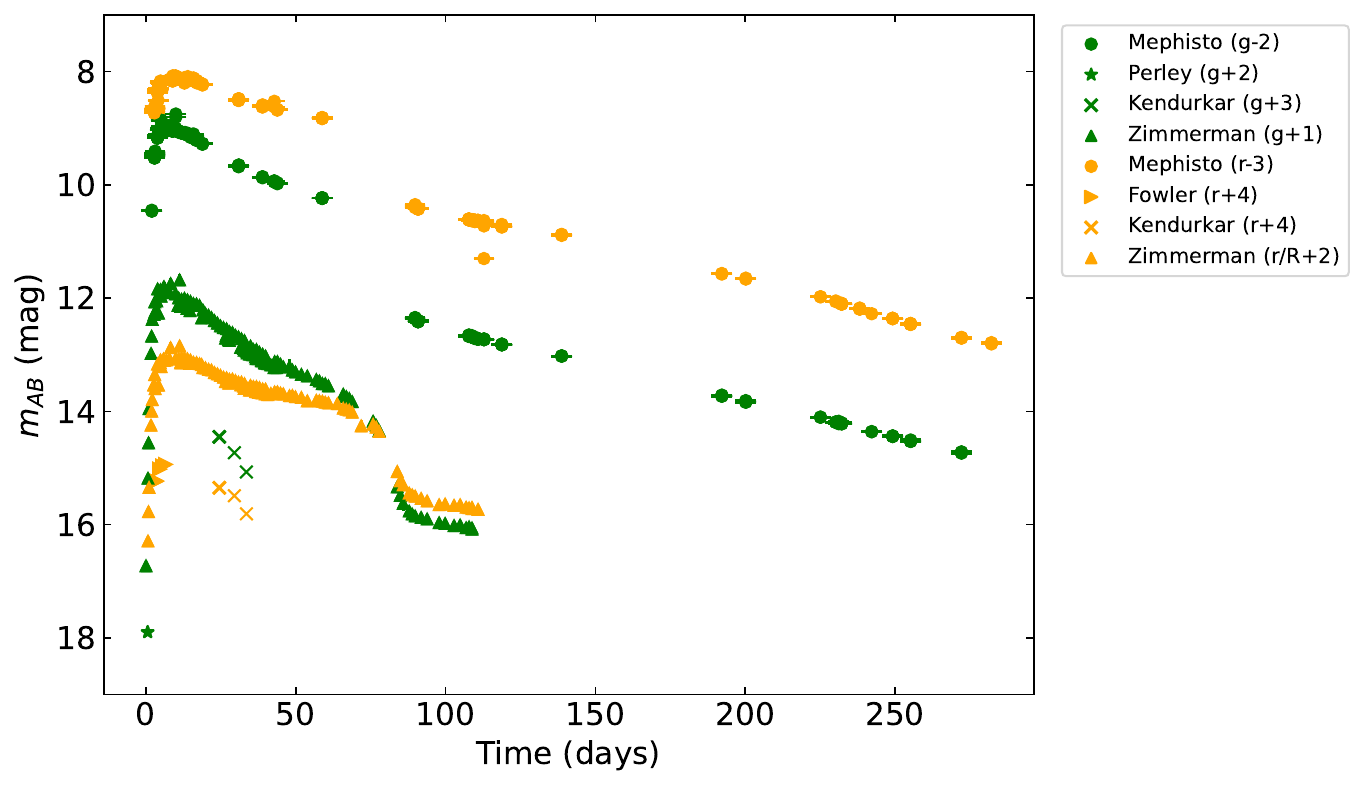}
    \caption{The comparison of light curves in $g$ (green points) and $r$ (yellow points) bands between this work and other previous works.
    The data from other previous works \citep{Perley23b,Kendurkar23d,Fowler23,Zimmerman23}. Offsets have been added to the magnitudes of different bands for clarity, as specified in the legend.}\label{data_lc1}
\end{figure} 

\begin{figure}
    \centering
    \includegraphics[width = 0.9\linewidth, trim = 0 0 30 30, clip]{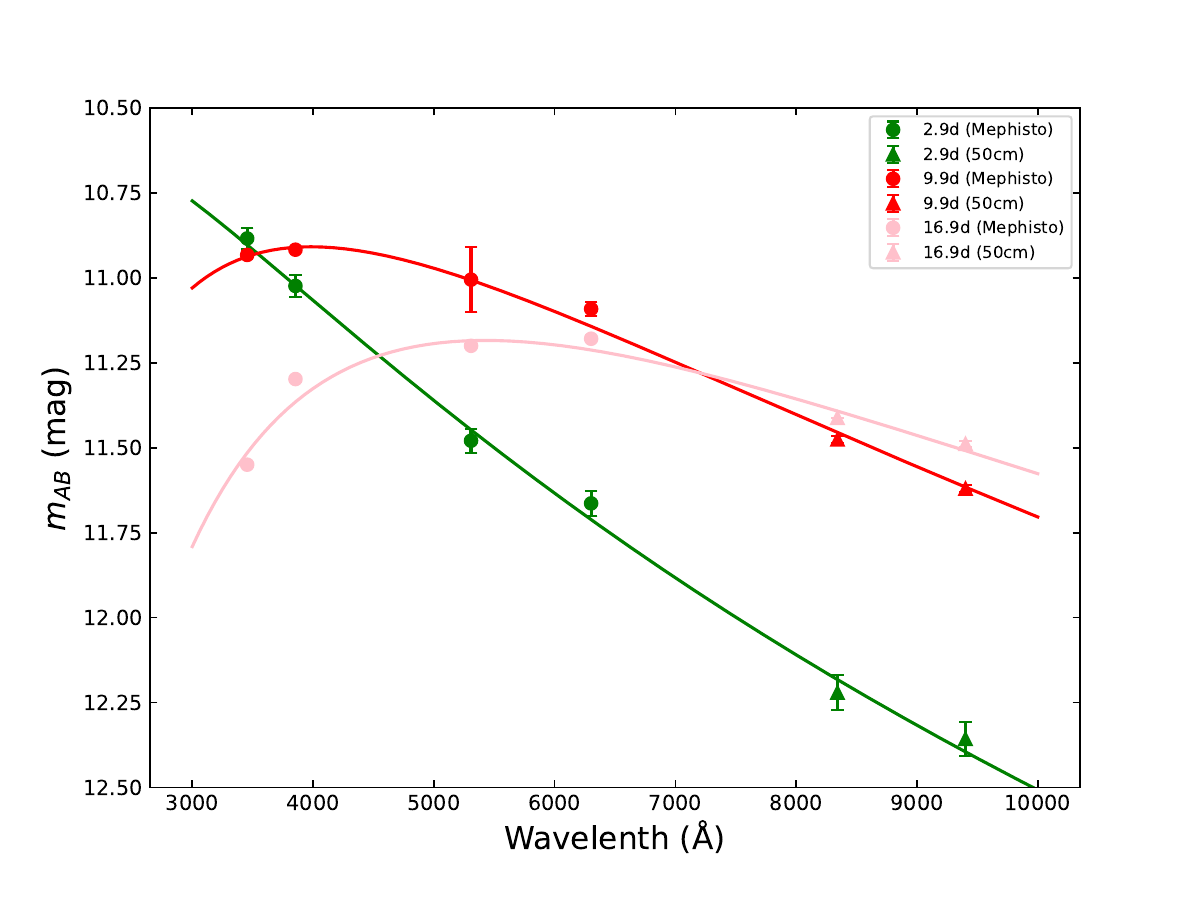}
    \caption{Spectral energy distributions (SEDs) at 2.9, 9.9, 16.9 days after phase zero. Circles and triangles correspond to apparent magnitudes measured with the Mephisto and the twin 50-cm telescopes, respectively. The green, red, and pink lines correspond to best-fit blackbody spectra of effective temperatures $T=25000,13000,9400~{\rm K}$, respectively. The wavelength of the data in this figure is taken as the mean wavelength of the filter.}\label{data_sed}
\end{figure}

In Figure \ref{data_lc}, we present the observed light curves of SN 2023ixf with Mephisto and the twin 50-cm telescopes in $uvgriz$ bands (the data behind Figure \ref{data_lc} are shown in the appendix, and we should notice that the uncertainties in the appendix mainly include the photon-counting uncertainties. The uncertainties of the photometric calibration and the zero points of magnitude are approximately $0.01~{\rm mag}$). All photometric data are in AB magnitudes and have been corrected for the Milky Way and host galaxy extinctions. Since the measurement errors of Mephisto are much smaller than those of the twin 50-cm telescopes, for $gr$ bands, we only include the Mephisto measurements in this work.
The Mephisto measurements show that SN 2023ixf has peak magnitudes $m_u=10.59$ mag, $m_v=10.71$ mag, $m_g=11.14$ mag, and $m_r=11.33$ mag at $\sim3.9$ days after phase zero and decline gradually after the explosion. 
The rise time (post-explosion) of SN 2023ixf is significantly shorter than the average rise time of other normal Type II events \citep[c.f. $\sim$10 days, see][]{Valenti2016}.
The comparison of light curves in $g$ and $r$ bands between this work and other previous works \citep{Perley23b,Kendurkar23d,Fowler23,Zimmerman23} are shown in Figure \ref{data_lc1}.
In Figure \ref{data_sed}, we plot the spectral energy distribution (SED) at 2.9, 9.9, and 16.9 days after phase zero and fit them with a blackbody. 
The mean wavelengths of the $uvgriz$ bands are $3449.95$, $3888.79$, $5273.22$, $6253.80$, $8301.75$, and $9368.17$\AA, respectively.
One can see that the effective temperature evolves significantly during the early phase of SN 2023ixf.

\section{Evolution of Bolometric Luminosity, Effective Temperature and Photospheric Radius}\label{fireball}

In this section, we discuss the evolving properties of SN 2023ixf.
The evolution of an SN could be well described by an isotropic expanding fireball in the early phase. For an SN ejecta with a time-dependent mean density $\rho(t)$, the evolution of the photospheric radius could be described by \citep[e.g.,][]{Arnett82,Liu18},

\begin{align}
R_{\rm ph}(t)=R(t)-\frac{2}{3}\lambda(t),
\end{align}
where $\lambda(t)=1/[\kappa(t)\rho(t)]$ the mean free path of photons and $\kappa$ the opacity of the SN ejecta. We should notice that the opacity $\kappa$ of Type II SNe could evolve due to the hydrogen recombination process.
The SN ejecta enters a homologous expansion phase after a few times of
the expansion timescale $R_{0}/v$, where $v$ is the ejecta velocity and $R_{0}$ the progenitor radius. At the early phase of SN expansion, the mean free path $\lambda(t)$ is much smaller than the ejecta radius $R(t)$, resulting in a photosphere that is homologously expanding, $R_{\rm ph}(t)\simeq R(t)$. As the ejecta density decreases with the expansion, the photospheric radius reaches its maximum at the transitional timescale $t_{\rm tr}$ with $dR_{\rm ph}(t_{\rm tr})/dt=0$ \citep[e.g.,][]{Liu18}. The photosphere then enters a late declining phase at $t>t_{\rm tr}$. 

\begin{figure}
    \centering
    \includegraphics[width = 0.9\linewidth, trim = 30 40 50 50, clip]{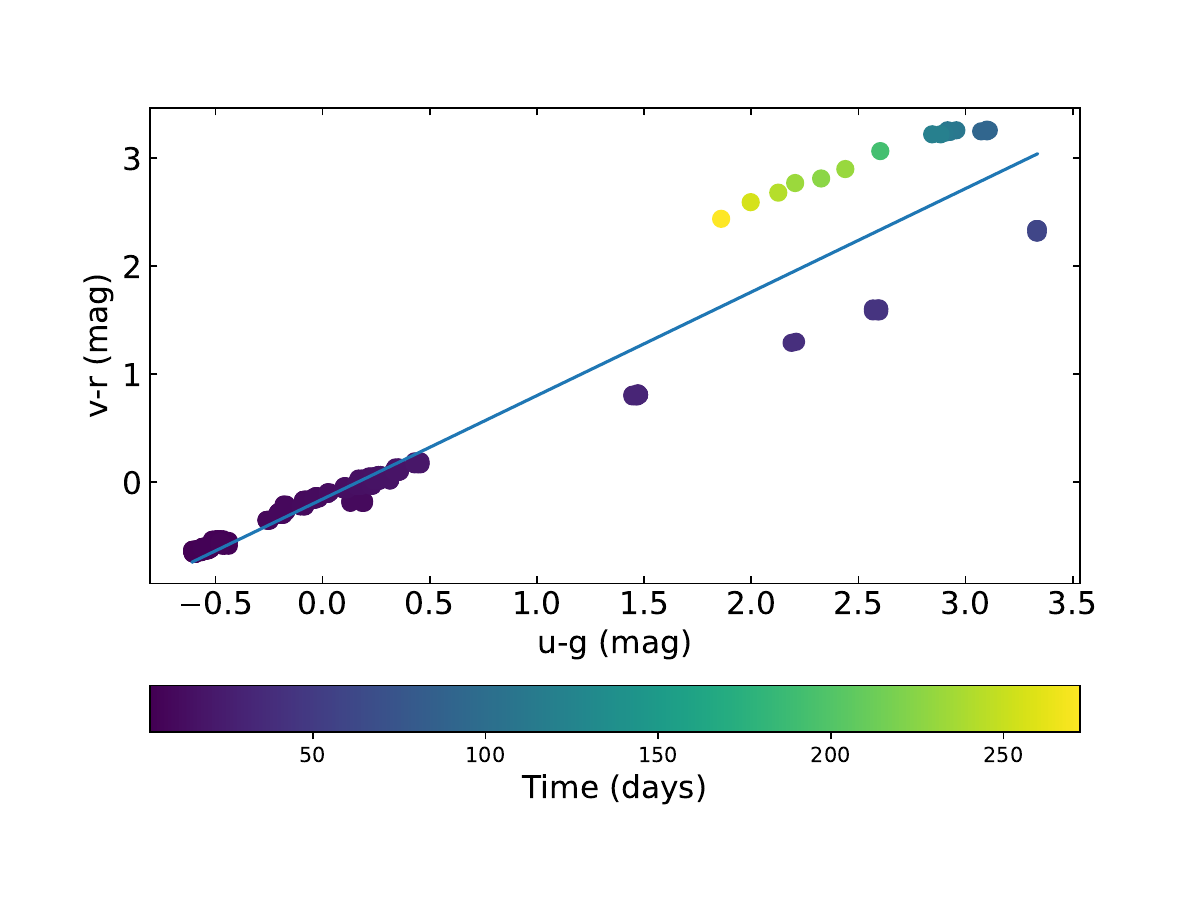}
    \caption{The evolution of SN 2023ixf in the $(u-g)-(v-r)$ color-color diagram. The epoch of the measurement after phase zero with Mephisto is represented by the color of the data point. 
    The solid line corresponds to the theoretical prediction of blackbody radiation based on Eq.(\ref{color}).}\label{c2c}
\end{figure}

Assuming that the blackbody radiation dominates the SN emission, we can derive the bolometric luminosity from the multiband photometry at each epoch and infer the effective temperature by fitting the SED at the same epoch.
First, the AB magnitude system is defined based on the observed flux density,

\begin{align}
m_{\rm AB,\nu}=-2.5\log F_\nu-48.6.
\end{align}
For a fireball of uniform brightness of a photospheric radius $R_{\rm ph}$ and a radiation intensity $B_\nu$, the flux density at distance $d$ is $F_\nu=\pi B_\nu(R_{\rm ph}/d)^2$, where the intensity of the blackbody radiation satisfies the Planck law,

\begin{align}
B_\nu(T_{\rm eff},\nu)=\frac{2h\nu^3}{c^2}\frac{1}{\exp(h\nu/kT_{\rm eff})-1},
\end{align}
where $T_{\rm eff}$ is the effective temperature of the SN ejecta.
We consider that the photometry is performed in two bands denoted by $i=1,2$ and the measured AB magnitude in a band of frequency $\nu_i$ is $m_{{\rm AB},\nu_i}$ ($\nu_1<\nu_2$). The color between band 1 and band 2 is,

\begin{align}
\Delta m_{\rm AB,\nu_{21}}= m_{\rm AB,\nu_2}-m_{\rm AB,\nu_1}=-2.5\log\left[\frac{B_\nu(T_{\rm eff},\nu_2)}{B_\nu(T_{\rm eff},\nu_1)}\right].\label{color}
\end{align}
The observed evolution of SN 2023ixf in the $(u-g)-(v-r)$ color-color diagram is shown in Figure \ref{c2c}. It indicates that the radiation is well approximated by blackbody radiation with some deviations at the later epochs. 
It is interesting that the observed color-color diagram in the later epochs is distributed on both sides of the theoretical prediction of blackbody radiation with a turning point occurring at $\sim90$ days. Such a turning feature is attributed to the effective temperature slightly rising after 90 days, see the following discussion.

Let $L_{\nu}$ be the specific luminosity at frequency $\nu$. According to the Planck law, the bolometric luminosity $L$ is given by,

\begin{align}
L=\frac{\pi^4}{15}\left(\frac{kT_{\rm eff}}{h\nu}\right)^4\left[\exp\left(\frac{h\nu}{kT_{\rm eff}}\right)-1\right]\nu L_{\nu}.
\end{align}
For $h\nu\ll kT_{\rm eff}$, one approximately has
$L\simeq(\pi^4/15)(kT_{\rm eff}/h\nu)^3\nu L_{\nu}$.
Using $F_\nu=L_\nu/4\pi d^2$, the AB magnitude at frequency $\nu$ could be written as,

\begin{align}
m_{\rm AB,\nu}=-2.5\log\left[\frac{15}{4\pi^5}\frac{L}{d^2\nu}\frac{(h\nu/kT_{\rm eff})^4}{\exp(h\nu/kT_{\rm eff})-1}\right]-48.6.
\end{align}
For a given SED (i.e., the apparent magnitude as a function of frequency) at time $t$, the bolometric luminosity $L(t)$ and the effective temperature $T_{\rm eff}(t)$ could be fitted based on the above relation once the distance $d$ is specified.
Furthermore, for the blackbody radiation, the photospheric radius at time $t$ can be derived by,

\begin{align}
R_{\rm ph}(t)=\sqrt{\frac{L(t)}{4\pi \sigma T_{\rm eff}(t)^4}},
\end{align}
where $\sigma=2\pi^5k^4/15c^2h^3=5.67\times10^{-5}~{\rm erg~cm^{-1}s^{-1}K^{-4}}$ the Stefan-Boltzmann constant. At last, we should notice that for broad filters as shown in Figure \ref{transmission}, accounting for the different widths of each filter, it would be more appropriate to convolve the passbands of the filters with the SED of the blackbody radiation instead of using a certain frequency.

\begin{figure} 
    \centering
    \includegraphics[width = 0.9\linewidth, trim = 0 0 0 0, clip]{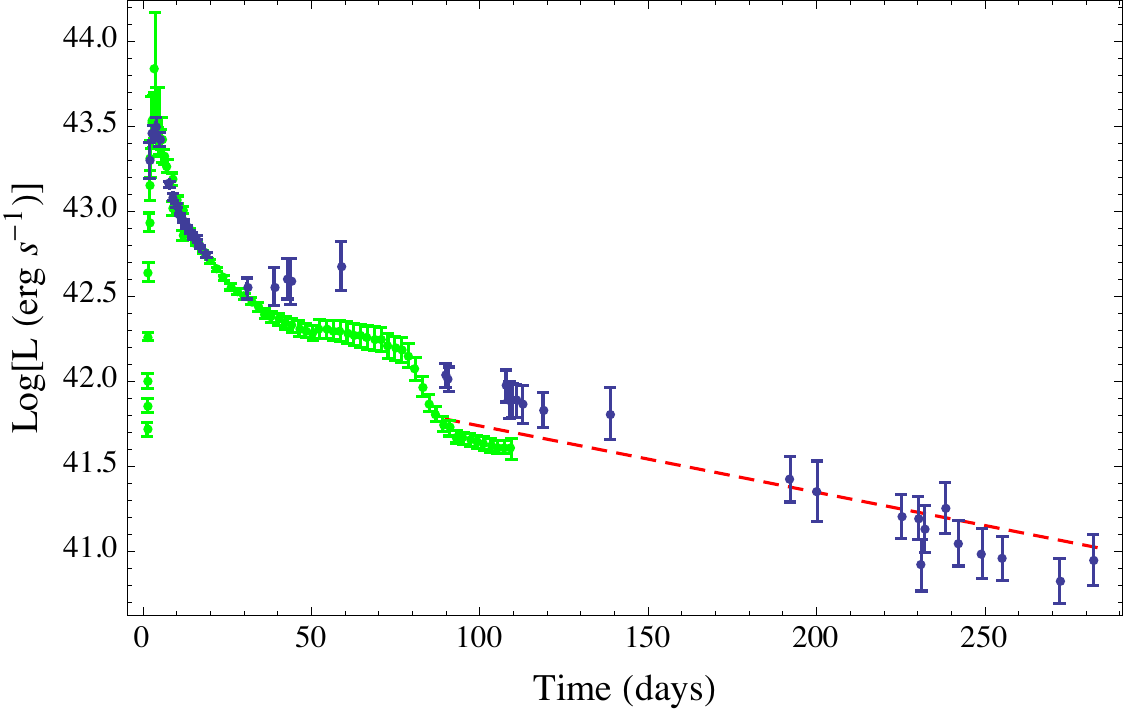}
    \includegraphics[width = 0.9\linewidth, trim = 0 0 0 0, clip]{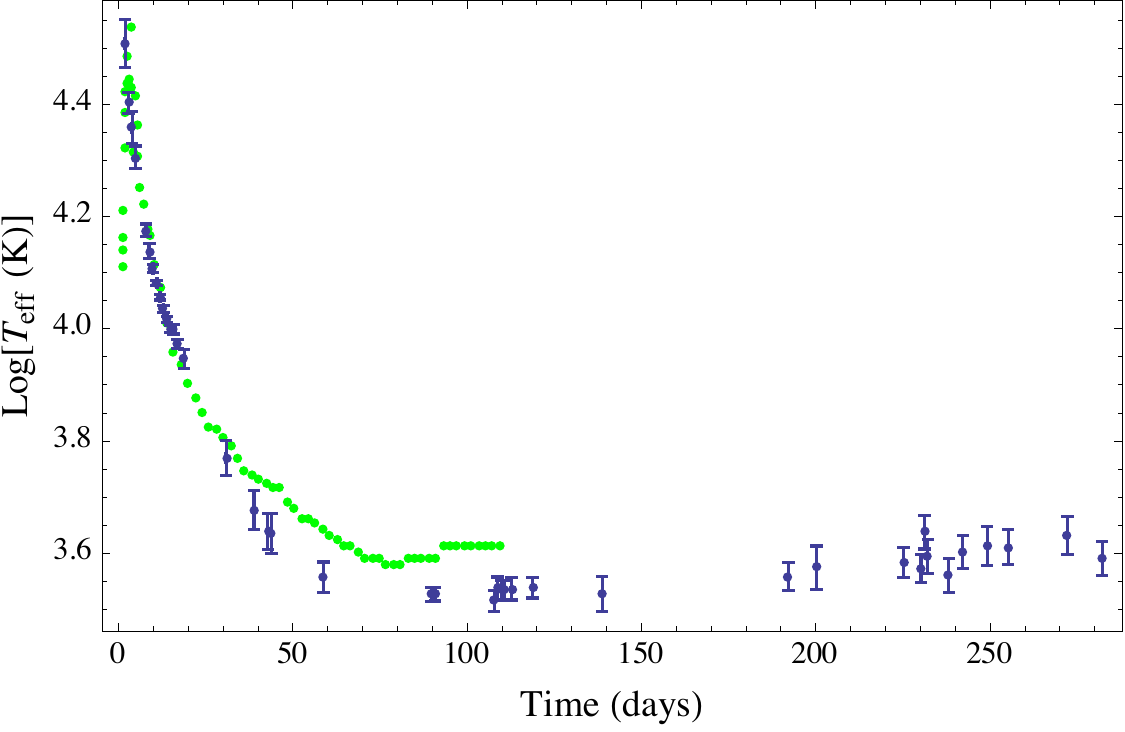}
    \includegraphics[width = 0.9\linewidth, trim = 0 0 0 0, clip]{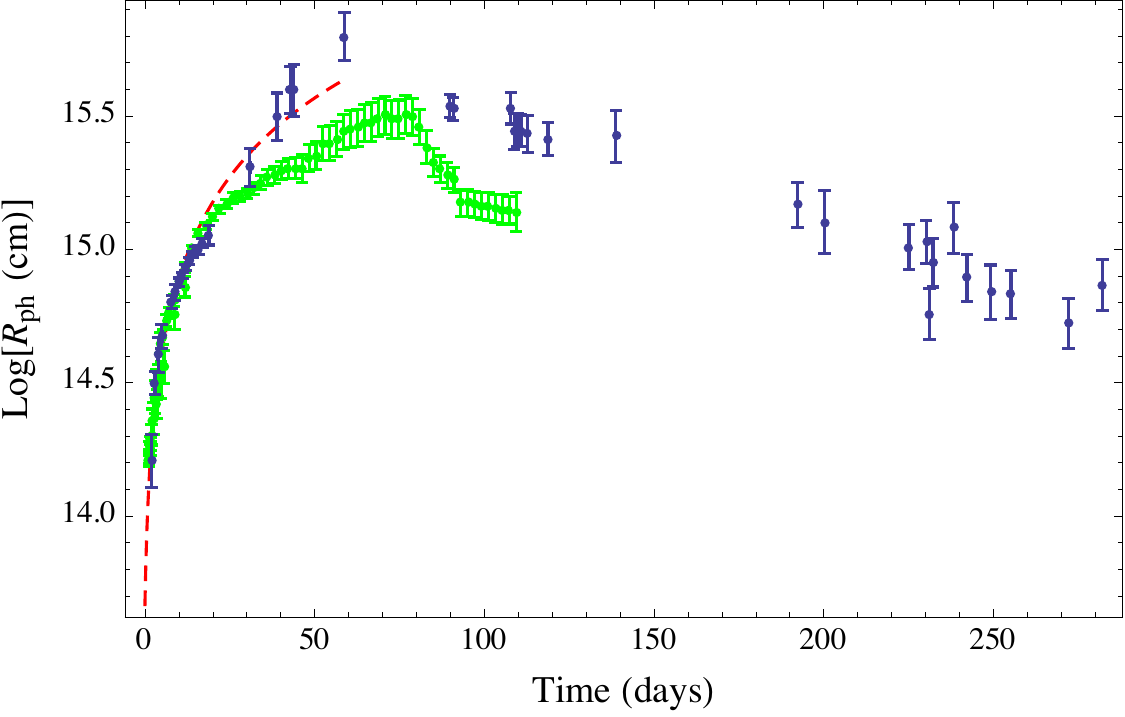}
    \caption{The evolution of bolometric luminosity (top panel), effective temperature (middle panel), and photospheric radius of SN 2023ixf. The blue points correspond to the data measured with Mephisto and the twin 50-cm telescopes, as shown in Table \ref{tabevo} of the Appendix. The green points correspond to the data from \citet{Zimmerman23}, which are shown for comparison. The red dashed line in the bolometric luminosity evolution (top panel) indicates the energy deposition from the best-fit $^{56}$Ni mass after the plateau. The red dashed line in the photospheric radius evolution (bottom panel) indicates the homologous expansion of the photosphere before hydrogen recombination. }\label{lc_tem}
\end{figure}

Based on the expanding fireball model, the evolution of bolometric luminosity (top panel), effective temperature (middle panel), and photospheric radius (bottom panel) are shown in Figure \ref{lc_tem} with the corresponding data in Table \ref{tabevo} of the Appendix. In doing so we used the data of the Mephisto and the twin 50-cm telescopes and combined the measurements on the same date and calculated the mean value and the standard deviation. We find that the bolometric luminosity reached a maximum value of $L\sim~3\times10^{43}~{\rm erg~s^{-1}}$ at 3.9 days after phase zero, which is $\sim45\%$ of the maximum luminosity given by \citet{Zimmerman23}. 
Such a deviation might be due to the following reasons: 1) the maximum luminosity of \citet{Zimmerman23} has a relatively large uncertainty, i.e., $\log(L/10^{42}~{\rm erg~s^{-1}})=68.59\pm78.18$, see the Supplementary Table 8 of \citet{Zimmerman23}. The maximum luminosity measured in this work is actually in the uncertainty range of \citet{Zimmerman23}.
2) \citet{Zimmerman23} used UV filters from Swift, while the Mephisto dataset went as blue as the $u$ band. The different observed band coverages between the two facilities cause a systematic deviation of the temperature measurement \citep[also see][as an example showing this issue]{Faran18}, finally leading to the bolometric luminosity deviation.
The bolometric luminosity evolution shows a plateau phase in the first 90 days after the explosion. After that, the light curve fully settled onto the radioactive tail. This feature implies a hydrogen recombination process in the first 90 days, which is also consistent with the observation of \citet{Zimmerman23}. 
We find that the measured bolometric luminosity in the first month is well consistent with that of \citet{Zimmerman23} but becomes larger than that of \citet{Zimmerman23} after the first month, meanwhile, for our measurement, the uncertainties of the bolometric luminosity after the first month are larger than those in the first month. The main reason is that the peak wavelength of the blackbody radiation has significantly shifted out from our observed bands after the first month, leading to a systematic deviation and large measurement uncertainties of the effective temperature and the bolometric luminosity, also see the SED evolution of Figure \ref{data_sed}.
The effective temperature shows a continuous decline in the first two months and gradually tends towards a constant in the following observation, as shown in the middle panel of Figure \ref{lc_tem}. The maximum effective temperature reaches $3.2\times10^4~{\rm K}$ at the first observation epoch and tends towards $\sim(3000-4000)~{\rm K}$ after the first two months. We also notice that the effective temperature slightly rises after 90 days (consistent with the turning feature in the color-color diagram, see Figure \ref{c2c}), although the data has relatively large uncertainties during this period. 
Similar to the evolution of the bolometric luminosity, due to the limit observed band range, the measured effected temperature in the first month is consistent with that of \citet{Zimmerman23} but becomes lower than that of \citet{Zimmerman23} after the first month. The evolution of photospheric radius presents a homologous expansion in the first two months and significantly shrinks after that.
The photospheric radius in the homologous expansion phase is given by,

\begin{align}
R_{\rm ph}(t)=R_\ast+v_{\rm ej}(t-t_\ast),\label{motion}
\end{align}
for $t>t_\ast$,
where $R_\ast$ and $t_\ast$ are the initial radius and time for homologous expansion, respectively, and $v_{\rm ej}$ is the ejecta velocity. 
Based on the above equation, we fit the evolution of the photospheric radius in the expansion phase. The best-fit result is shown in the red dashed line in the bottom panel of Figure \ref{lc_tem} and the photospheric evolution is consistent with a homologous expansion with

\begin{align} 
R_\ast\simeq722R_\odot,~t_\ast\simeq0.32~{\rm days},~v_{\rm ej}\simeq8.7\times10^3~{\rm km~s^{-1}}.
\end{align} 
The photospheric velocity is consistent with previous works \citep[e.g.,][]{Zimmerman23}, but the initial radius is smaller than that of \citet{Zimmerman23} that involved the constant-radius value as the shock-breakout radius of the SN evolution.

\section{Physical Properties of SN 2023ixf}\label{model}

In this section, we constrain the physical properties (e.g., the ejected mass $M_{\rm ej}$, explosion energy $E$, initial $^{56}$Ni mass, etc.) of SN 2023ixf based on the observed multi-wavelength light curves. 
For Type II SN, once the outer layers of ejecta cool below $T_{i}\simeq6000~{\rm K}$, the hydrogen recombination process will occur and the electron scattering opacity will drop by several orders of magnitude. In this case, the neutral material above the ionization front is transparent, and ionized material inside the front is opaque, i.e., $\kappa\sim{\rm const}$ for $T>T_i$ and $\kappa\sim0$ for $T<T_i$, and the photosphere is nearly coincident with the ionization front. As radiation escapes, the ionization front recedes inward in the ejecta's comoving frame. 
When the ionization front reaches the base of the hydrogen envelope, the internal energy will be largely depleted, leading to the end of the plateau where the light curve drops off sharply.
We find that the evolution of the bolometric luminosity, effective temperature, and photospheric radius support the above physical picture: 
the light curve shows a plateau phase in the first 90 days and fully settles onto the radioactive tail subsequently, meanwhile, the effective temperature tends towards a constant and the photospheric radius shrinks at $\sim(50-60)~{\rm days}$ after the explosion.
An approximate analytic scaling rule for Type II SNe has been derived for both the luminosity $L_p$ and the duration $\tau_p$ of the plateau \citep{Popov93,Kasen09,Sukhbold16}. 
Calibrating the analytic scaling rule to a set of numerical models, the luminosity and duration of the plateau are given by \citep{Kasen09,Sukhbold16},

\begin{align}
&L_{p,50}\simeq1.26\times10^{42}~{\rm erg~s^{-1}} E_{51}^{5/6}M_{\rm ej,10}^{-1/2}R_{0,500}^{2/3},\label{lp}\\
&t_p\simeq122~{\rm day} E_{51}^{-1/4}M_{\rm ej,10}^{1/2}R_{0,500}^{1/6},\label{tp}
\end{align}
where $M_{\rm ej,10}=M_{\rm ej}/10M_\odot$, $R_{0,500}=R_0/500R_\odot$, $L_{p,50}$ is the plateau luminosity at 50 days, and the helium mass fraction is assumed to be $X_{\rm He}=0.33$ here. Thus, the explosion energy $E$ and the ejecta mass $M_{\rm ej}$ could be estimated by the above approximate analytic scaling rules.
In this work, the plateau luminosity of SN 2023ixf at 50 days is measured to be $L_{p,50}\simeq4\times10^{42}~{\rm erg~s^{-1}}$ (see the top panel of Figure \ref{lc_tem}), and the plateau duration is estimated to be $t_p\simeq90~{\rm day}$ \citep[also see][]{Zimmerman23}. The progenitor radius is taken as $R_0\simeq(410-1400)R_\odot$ based on the bolometric luminosity and the effective temperature of the progenitor \citep{Niu23,Hosseinzadeh23,Jencson23,Kilpatrick23,Soraisam23,Xiang23,Qin23}. Using Eq.(\ref{lp}) and Eq.(\ref{tp}), the explosion energy and the ejecta mass are constrained to be,

\begin{align}
&E\simeq (1.0-5.7)\times10^{51}~{\rm erg},\\
&M_{\rm ej}\simeq(3.8-13.9)M_\odot.
\end{align}
We notice that the uncertainties of $E$ and $M_{\rm ej}$ are mainly attributed to the different measurements of the progenitor radius $R_0$ in the literature.

In addition to the method that uses the scaling rule between $L_p$ and $t_p$, the ejecta mass can also be measured by the time-weighted integrated luminosity removing the contribution from the initial $^{56}$Ni. 
Since there is a clear distinction at 90 days between the photospheric phase and the radioactive tail in the light curve of SN 2023ixf, as shown in the top panel of Figure \ref{lc_tem} \citep[also see][]{Zimmerman23} the contributions of the cooling envelope and $^{56}$Ni decay to the photospheric emission could be separated based on the measurement of the time-weighted integrated bolometric luminosity \citep{Katz13,Nakar16}. 
We define the internal energy as $E_{\rm int}(t)$ and the input power by the radioactive decay as $Q_{\rm Ni}(t)$.
For a homologously expanding ejecta, the derivative of the internal energy is,

\begin{align}
\frac{dE_{\rm int}(t)}{dt}=-\frac{E_{\rm int}(t)}{t}+Q_{\rm Ni}(t)-L(t),\label{dEdt}
\end{align} 
where $E_{\rm int}(t)/t$ corresponds to the loss rate of the internal energy due to the adiabatic expansion, $L(t)$ the bolometric luminosity. The input power by the radioactive decay is given by\citep{Arnett82,Chatzopoulos12},

\begin{align}
Q_{\rm Ni}(t)=M_{\rm Ni}\left[(\epsilon_{\rm Ni}-\epsilon_{\rm Co})e^{-t/t_{\rm Ni}}+\epsilon_{\rm Co}e^{-t/t_{\rm Co}}\right],\label{Qni}
\end{align}
where $M_{\rm Ni}$ the initial nickel mass of the SN ejecta, $t_{\rm Ni}=8.8~{\rm days}$, $t_{\rm Co}=111.3~{\rm days}$, $\epsilon_{\rm Ni}=3.9\times10^{10}~{\rm erg~s^{-1}g^{-1}}$ and $\epsilon_{\rm Co}=6.8\times10^9~{\rm erg~s^{-1}g^{-1}}$ are the energy generation rates due to Ni and Co decays, respectively. At $t>t_p$, the diffusion time through the envelope is much shorter than $t$, leading to $L(t>t_p)=Q_{\rm Ni}(t>t_p)$. Therefore, the initial $^{56}$Ni mass can be measured by the observed radioactive tail after 90 days. In the top panel of Figure \ref{lc_tem}, we fit the bolometric luminosity after 90 days using Eq.(\ref{Qni}) and obtain an initial $^{56}$Ni mass,

\begin{align}
M_{\rm Ni}\simeq0.098M_\odot,
\end{align}
which is 27\% larger than $M_{\rm Ni}\simeq0.071M_\odot$ given by \citet{Zimmerman23}.
Using the partial integration for Eq.(\ref{dEdt}), one has \citep{Katz13,Nakar16},

\begin{align}
ET\equiv E_{\rm int}(t_\ast)t_\ast=\int_0^{t_p}t[L(t)-Q_{\rm Ni}(t)]dt,
\end{align}
where $t_\ast$ the time at which the homologous expansion phase begins \citep{Nakar16}. Since the initial $^{56}$Ni mass has been measured by the radioactive tail, i.e., $M_{\rm Ni}=0.098M_\odot$. Using the bolometric luminosity evolution $L(t)$ and taking $t_p=90~{\rm day}$, we obtain

\begin{align}
ET\simeq7.7\times10^{55}~{\rm erg~s}.
\end{align}
On the other hand, as an observable quantity, $ET$ satisfies $ET\propto E R_0/v_{\rm ej}\propto M_{\rm ej}R_0v_{\rm ej}$, which depends on $M_{\rm ej}$, $R_0$, and $v_{\rm ej}$ and is insensitive to the exact details of radiation transport and CSM mass. The simulation of a large set of SN progenitors suggests a scaling relation of \citep{Shussman16}

\begin{align}
ET\simeq0.1M_{\rm ej}R_0v_{\rm ej}.
\end{align}
The progenitor radius has been measured to be $R_0\simeq(410-1400)R_\odot$ based on the bolometric luminosity and the effective temperature of the progenitor \citep{Niu23,Hosseinzadeh23,Jencson23,Kilpatrick23,Soraisam23,Xiang23,Qin23}. The ejecta velocity is measured to be $v_{\rm ej}\simeq8.7\times10^3~{\rm km~s^{-1}}$ based on the evolution of the photospheric radius in the early stage. Thus, the ejecta mass is estimated to be

\begin{align}
M_{\rm ej}\simeq\frac{ET}{0.1R_0v_{\rm ej}}\simeq(4.7-16)M_\odot,
\end{align}
which is approximately consistent with $M_{\rm ej}\simeq(3.8-13.9)M_\odot$ given by the scaling rule.

The peak bolometric luminosity of SN 2023ixf is $L_0\simeq3\times10^{43}~{\rm erg~s^{-1}}$ at $t_0\simeq3.9~{\rm days}$ after the explosion, which has been proposed to be caused by the interaction between the SN ejecta and the circumstellar medium (CSM). Assuming that the CMS density is $\rho_{\rm CSM}$ and the peak bolometric luminosity $L_0$ corresponds to the kinetic energy of the shock caused by the ejecta-CSM interaction, $L_0t_0\sim(1/2)M_{\rm CSM} v_{\rm ej}^2$, where the shock medium is dominated by the swept CSM,
one has the mass of the shocked CSM of

\begin{align}
M_{\rm CSM}\sim\frac{2L_0t_0}{v_{\rm ej}^2}\simeq0.013M_\odot.
\end{align}
The CSM density is estimated by $M_{\rm CSM}\sim(4\pi/3)(v_{\rm ej}t_0)^3\rho_{\rm CSM}$, leading to

\begin{align}
\rho_{\rm CSM}\sim\frac{3L_0}{2\pi v_{\rm ej}^5t_0^2}\simeq2.5\times10^{-13}~{\rm g~cm^{-3}}L_{0,{43.5}}t_{\rm 0, 3.9 day}^{-2},
\end{align}
where $t_{\rm 0, 3.9 day}=t_{0}/(3.9~{\rm day})$ and the ejecta velocity is taken as $v_{\rm ej}\simeq8700~{\rm km~s^{-1}}$. 
The mass loss rate of the stellar wind generating the CSM by the progenitor is,

\begin{align}
\dot M\sim 4\pi (v_{\rm ej}t_0)^2\rho_{\rm CSM}v_w\simeq0.022M_\odot~{\rm yr^{-1}}L_{0,{43.5}},
\end{align}
where the progenitor wind velocity is assumed to be $v_w=50~{\rm km~s^{-1}}$ as the typical value. The above CMS density and the mass loss rate of the progenitor are consistent with the results of \citet{Jacobson-Galan23} and \citet{Zimmerman23}.
There are some possible ways to form the CSM near the SN progenitor as proposed by \citet{Branch17}: 1) the fast wind of the massive progenitor produces low-density bubbles surrounded by the swept-up interstellar matter. 2) an episodic matter ejection from the progenitor (or from a common-envelope episode of a binary system) forms the detached CSM shell. 3) the fast wind of the progenitor that has returned to the blue compresses the previous red-supergiant wind into a shell.

\section{Discussions and Conclusions}\label{conclusion}

In this paper, we presented the multi-wavelength simultaneous photometric observation of SN 2023ixf located in the nearby spiral host galaxy M101 at $d=6.85~{\rm Mpc}$ in the first 283 days after the explosion. 
Based on the measurements of the Mephisto and the twin 50-cm telescopes, we calculated and analyzed the evolution of the bolometric luminosity, effective temperature, and photospheric radius. 
The following conclusions for SN 2023ixf are drawn:

The measurements with the Mephisto showed that the peak apparent magnitudes of SN 2023ixf were $m_u=10.59$, $m_v=10.71$, $m_g=11.14$ and $m_r=11.33$ at 3.9 days after phase zero and the brightness declined in the following observation. We find that there is a significant drop in the light curve between 59 days and 90 days, which is consistent with the plateau end reported by \citet{Zimmerman23}.
Based on the multi-wavelength photometric data, we derived the evolution of the bolometric luminosity, effective temperature, and photospheric radius. 
The bolometric luminosity reached a maximum value of $L\sim~3\times10^{43}~{\rm erg~s^{-1}}$ at 3.9 days after phase zero and fully settled onto the radiative tail at $\sim90~{\rm days}$.
The effective temperature has been a maximum ($\gtrsim3.2\times10^4~{\rm K}$) at the first observation epochs and approached to a constant of $(3000-4000)~{\rm K}$ from 60 days to 283 days. 
The early-phase photospheric radius evolution is consistent with a homologous expansion with an initial radius of $R_\ast\simeq722R_\odot$ and an ejecta velocity of $v_{\rm ej}\simeq8.7\times10^3~{\rm km~s^{-1}}$ and gradually shrunk 60 days after the explosion. The above features of the bolometric luminosity, effective temperature, and photospheric radius suggest that SN 2023ixf represented a significant hydrogen recombination process in the first three months.

Since the light curve showed a significant radiative tail after $\sim90~{\rm days}$, we estimated the initial nickel mass in the SN ejecta as $M_{\rm Ni}\sim 0.098M_\odot$, which is 27\% larger than the result of \citet{Zimmerman23}. 
We first estimated the explosion energy $E$ and the ejecta mass $M_{\rm ej}$ based on the scaling rule of Type II SNe given by \citet{Kasen09} and obtained $E\simeq(1.0-5.7)\times10^{51}~{\rm erg}$ and $M_{\rm ej}\simeq(3.8-13.9)M_\odot$, respectively. 
On the other hand, due to the existence of the radiative tail in the light curve of SN 2023ixf, we can separate the contributions of the cooling envelope and $^{56}$Ni decay to the photospheric emission based on the measurement of the time-weighted integrated bolometric luminosity \citep{Katz13,Nakar16}. In this case, the mass of the SN ejecta is estimated to be $M_{\rm ej}\sim (4.7-16)M_\odot$.
We propose that the peak bolometric luminosity is caused by the interaction between the ejecta and the circumstellar medium (CSM). In this scenario, the shocked CSM mass is $M_{\rm CSM}\sim0.013M_\odot$, the CSM density is $\rho_{\rm CSM}\sim2.5\times10^{-13}~{\rm g~cm^{-3}}$ and the mass loss rate of the progenitor is $\dot M\sim0.022M_\odot~{\rm yr^{-1}}$.

\acknowledgments

Mephisto is developed at and operated by the South-Western Institute for Astronomy Research of Yunnan University (SWIFAR-YNU), funded by the ``Yunnan University Development Plan for World-Class University'' and ``Yunnan University Development Plan for World-Class Astronomy Discipline''. The authors acknowledge supports from the ``Science \& Technology Champion Project'' (202005AB160002) and from two ``Team Projects'' -- the ``Innovation Team'' (202105AE160021) and the ``Top Team'' (202305AT350002), all funded by the ``Yunnan Revitalization Talent Support Program''.
Y.P.Y. is supported by the National Natural Science Foundation of China grant No.12003028 and the National SKA Program of China (2022SKA0130100). X.K.L. acknowledges the supports from NSFC of China under grant No. 12173033, National Key R\&D Program of China No. 2022YFF0503403, and YNU grant No. C176220100008. 
Y.-Z. Cai is supported by the National Natural Science Foundation of China (NSFC, Grant No. 12303054), the Yunnan Fundamental Research Projects (Grant No. 202401AU070063) and the International Centre of Supernovae, Yunnan Key Laboratory (No. 202302AN360001).
We thank the anonymous referee for providing helpful comments and suggestions that have allowed us to improve this manuscript significantly.
We also acknowledge the helpful discussions with Liang-Duan Liu, Jin-Ping Zhu, Weili Lin, Erez Zimmerman and Andrea Reguitti (INAF-OAPd).


\appendix

In this appendix, we provide the photometry data of Mephisto (Table \ref{tabm}) and the twin 50-cm telescopes (Table \ref{tabf}) as the data behind Figure \ref{data_lc}. The best-fit results of bolometric luminosity, effective temperature, and photospheric radius for blackbody radiation are shown in Table \ref{tabevo} as the data behind Figure \ref{lc_tem}.

\begin{longtable}{ccccc}
  \caption{Photometry data of SN 2023ixf with Mephisto in $ug$ and $vr$ bands. This table is available in its entirety in the online supplementary material, and the full table can also be found at the ScienceDB \url{https://doi.org/10.57760/sciencedb.15753}. Notice that the uncertainties in the table mainly include the photon-counting uncertainties. The uncertainties of the photometric calibration and the zero points of magnitude are approximately $0.01~{\rm mag}$.}\label{tabm} \\
  \toprule
  \textbf{MJD/day}  & \textbf{u/mag} & \textbf{g/mag} & \textbf{v/mag} & \textbf{r/mag} \\
  \midrule
  \endfirsthead
  \multicolumn{5}{c}{{\tablename\ \thetable{} -- continued}} \\
  \toprule
 \textbf{MJD/day}  & \textbf{u/mag} & \textbf{g/mag} & \textbf{v/mag}& \textbf{r/mag}\\
  \midrule
  \endhead
  \bottomrule
  \endfoot
  \bottomrule
  \endlastfoot
  
60084.70538&--&12.448${\pm}$0.001&--&--\\
60084.70539&11.858${\pm}$0.001&--&--&--\\
60084.70627&--&12.463${\pm}$0.001&--&--\\
60084.70628&11.866${\pm}$0.001&--&--&--\\
60084.70735&--&12.463${\pm}$0.001&--&--\\
60084.70736&11.858${\pm}$0.001&--&--&--\\
60084.70846&11.867${\pm}$0.001&--&--&--\\
60084.70932&11.863${\pm}$0.001&--&--&--\\
60084.71094&--&--&12.004${\pm}$0.002&--\\
60084.71179&--&--&12.008${\pm}$0.002&--\\
...&...&...&...&...\\
60354.93491&--&--&--&15.702${\pm}$0.008\\
60354.93715&--&--&--&15.695${\pm}$0.008\\
60354.93939&--&--&18.14${\pm}$0.03&--\\
60354.93941&--&--&--&15.694${\pm}$0.008\\
60354.94166&--&--&--&15.706${\pm}$0.008\\
60364.93336&--&--&18.18${\pm}$0.03&15.786${\pm}$0.008\\
60364.93560&--&--&--&15.809${\pm}$0.008\\
60364.93771&--&--&18.28${\pm}$0.03&--\\
60364.93785&--&--&--&15.796${\pm}$0.008\\
60364.94009&--&--&--&15.796${\pm}$0.008\\
\end{longtable}

\begin{longtable}{ccc}
  \caption{Photometry data of SN 2023ixf with the twin 50-cm telescopes in $iz$ band. This table is available in its entirety in the online supplementary material, and the full table can also be found at the ScienceDB \url{https://doi.org/10.57760/sciencedb.15753}. Notice that the uncertainties in the table mainly include the photon-counting uncertainties. The uncertainties of the photometric calibration and the zero points of magnitude are approximately $0.01~{\rm mag}$.} \label{tabf} \\
  \toprule
  \textbf{MJD/day}  & \textbf{i/mag} & \textbf{z/mag} \\
  \midrule
  \endfirsthead
  \multicolumn{3}{c}{{\tablename\ \thetable{} -- continued}} \\
  \toprule
 \textbf{MJD/day}  & \textbf{i/mag} & \textbf{z/mag} \\
  \midrule
  \endhead
  \bottomrule
  \endfoot
  \bottomrule
  \endlastfoot
60084.66164&13.423${\pm}$0.005&--\\
60084.66464&--&13.58${\pm}$0.01\\
60084.85436&13.067${\pm}$0.003&--\\
60085.65457&12.280${\pm}$0.002&--\\
60085.65757&12.291${\pm}$0.002&--\\
60085.65759&--&12.459${\pm}$0.006\\
60085.74509&12.220${\pm}$0.002&--\\
60085.74809&12.221${\pm}$0.002&--\\
60085.85156&12.156${\pm}$0.002&--\\
60085.85160&--&12.361${\pm}$0.006\\
...&...&...\\
60096.62810&11.383${\pm}$0.001&--\\
60096.62811&--&11.496${\pm}$0.003\\
60097.62223&11.395${\pm}$0.001&11.503${\pm}$0.003\\
60097.62383&11.394${\pm}$0.001&11.488${\pm}$0.003\\
60098.62184&11.393${\pm}$0.001&--\\
60098.62341&11.398${\pm}$0.001&--\\
60098.62344&--&11.503${\pm}$0.003\\
60099.61574&--&11.502${\pm}$0.003\\
60099.61575&11.412${\pm}$0.001&--\\
60099.61734&11.412${\pm}$0.001&11.483${\pm}$0.003\\
\end{longtable}

\begin{longtable}{cccc}
  \caption{The best-fit results of bolometric luminosity, effective temperature, and photospheric radius for blackbody radiation.} \label{tabevo} \\
  \toprule
  \textbf{MJD/day}  & \textbf{$\log[L/{\rm erg~s^{-1}}]$} & \textbf{$\log[T_{\rm eff}/{\rm K}]$} & \textbf{$\log[R_{\rm ph}/{\rm cm}]$} \\
  \midrule
  \endfirsthead
  \multicolumn{4}{c}{{\tablename\ \thetable{} -- continued}} \\
  \toprule
 \textbf{MJD/day}  & \textbf{$\log[L/{\rm erg~s^{-1}}]$} & \textbf{$\log[T_{\rm eff}/{\rm K}]$} & \textbf{$\log[R_{\rm ph}/{\rm cm}]$} \\
  \midrule
  \endhead
  \bottomrule
  \endfoot
  \bottomrule
  \endlastfoot
60084.70734		&	43.29	$\pm$	0.10	&	4.507	$\pm$	0.042	&	14.206	$\pm$	0.099	\\
60085.70783		&	43.45	$\pm$	0.04	&	4.401	$\pm$	0.018	&	14.498	$\pm$	0.044	\\
60086.67782		&	43.49	$\pm$	0.06	&	4.357	$\pm$	0.028	&	14.603	$\pm$	0.064	\\
60087.67758		&	43.41	$\pm$	0.04	&	4.305	$\pm$	0.020	&	14.673	$\pm$	0.045	\\
60090.65298		&	43.15	$\pm$	0.01	&	4.175	$\pm$	0.011	&	14.802	$\pm$	0.024	\\
60091.62564		&	43.08	$\pm$	0.02	&	4.137	$\pm$	0.013	&	14.838	$\pm$	0.029	\\
60092.71010		&	43.03	$\pm$	0.01	&	4.106	$\pm$	0.007	&	14.876	$\pm$	0.015	\\
60093.74334		&	42.98	$\pm$	0.01	&	4.080	$\pm$	0.004	&	14.901	$\pm$	0.009	\\
60094.71458		&	42.94	$\pm$	0.01	&	4.056	$\pm$	0.005	&	14.931	$\pm$	0.011	\\
60095.68900		&	42.90	$\pm$	0.01	&	4.035	$\pm$	0.005	&	14.955	$\pm$	0.012	\\
60096.69822		&	42.87	$\pm$	0.01	&	4.015	$\pm$	0.005	&	14.979	$\pm$	0.011	\\
60097.67836		&	42.84	$\pm$	0.01	&	3.999	$\pm$	0.006	&	14.996	$\pm$	0.014	\\
60098.59604		&	42.84	$\pm$	0.01	&	3.998	$\pm$	0.008	&	14.997	$\pm$	0.017	\\
60099.68435		&	42.78	$\pm$	0.01	&	3.972	$\pm$	0.007	&	15.023	$\pm$	0.016	\\
60101.61025		&	42.74	$\pm$	0.01	&	3.945	$\pm$	0.017	&	15.052	$\pm$	0.036	\\
60113.66383		&	42.54	$\pm$	0.06	&	3.769	$\pm$	0.031	&	15.306	$\pm$	0.070	\\
60121.66017		&	42.55	$\pm$	0.11	&	3.677	$\pm$	0.035	&	15.496	$\pm$	0.089	\\
60125.60418		&	42.60	$\pm$	0.12	&	3.638	$\pm$	0.031	&	15.597	$\pm$	0.087	\\
60126.59941		&	42.58	$\pm$	0.13	&	3.635	$\pm$	0.035	&	15.595	$\pm$	0.097	\\
60141.55753		&	42.67	$\pm$	0.14	&	3.557	$\pm$	0.027	&	15.797	$\pm$	0.090	\\
60172.56913		&	42.03	$\pm$	0.07	&	3.526	$\pm$	0.012	&	15.537	$\pm$	0.043	\\
60173.58373		&	42.01	$\pm$	0.07	&	3.526	$\pm$	0.012	&	15.525	$\pm$	0.043	\\
60190.53759		&	41.97	$\pm$	0.09	&	3.514	$\pm$	0.018	&	15.529	$\pm$	0.059	\\
60191.53887		&	41.88	$\pm$	0.10	&	3.537	$\pm$	0.020	&	15.442	$\pm$	0.067	\\
60192.53781		&	41.88	$\pm$	0.10	&	3.534	$\pm$	0.017	&	15.448	$\pm$	0.062	\\
60193.54465		&	41.88	$\pm$	0.09	&	3.534	$\pm$	0.017	&	15.445	$\pm$	0.060	\\
60195.56868		&	41.86	$\pm$	0.11	&	3.536	$\pm$	0.019	&	15.432	$\pm$	0.069	\\
60201.52122		&	41.83	$\pm$	0.10	&	3.538	$\pm$	0.017	&	15.413	$\pm$	0.062	\\
60221.49577		&	41.80	$\pm$	0.15	&	3.527	$\pm$	0.030	&	15.423	$\pm$	0.097	\\
60274.95250		&	41.42	$\pm$	0.13	&	3.559	$\pm$	0.025	&	15.166	$\pm$	0.083	\\
60282.94518		&	41.35	$\pm$	0.17	&	3.574	$\pm$	0.038	&	15.100	$\pm$	0.118	\\
60307.87612		&	41.20	$\pm$	0.12	&	3.583	$\pm$	0.026	&	15.009	$\pm$	0.083	\\
60312.95936		&	41.19	$\pm$	0.12	&	3.572	$\pm$	0.025	&	15.026	$\pm$	0.080	\\
60313.96008		&	40.91	$\pm$	0.15	&	3.637	$\pm$	0.030	&	14.756	$\pm$	0.096	\\
60314.93569		&	41.12	$\pm$	0.13	&	3.593	$\pm$	0.030	&	14.950	$\pm$	0.091	\\
60320.96748		&	41.25	$\pm$	0.15	&	3.560	$\pm$	0.030	&	15.080	$\pm$	0.096	\\
60324.96763		&	41.04	$\pm$	0.13	&	3.602	$\pm$	0.029	&	14.892	$\pm$	0.089	\\
60331.95188		&	40.98	$\pm$	0.14	&	3.613	$\pm$	0.034	&	14.839	$\pm$	0.101	\\
60337.95158		&	40.95	$\pm$	0.13	&	3.611	$\pm$	0.030	&	14.829	$\pm$	0.089	\\
60354.93162		&	40.82	$\pm$	0.13	&	3.632	$\pm$	0.033	&	14.722	$\pm$	0.094	\\
60364.93665		&	40.94	$\pm$	0.15	&	3.590	$\pm$	0.030	&	14.867	$\pm$	0.096	\\
\end{longtable}

\end{document}